\documentclass[useAMS,usenatbib]{mn2e}
\usepackage{graphicx,amssym}
\newcommand{\bc}{\begin{center}}
\newcommand{\ec}{\end{center}}

\title[High Redshift Galaxy Populations and their Descendants]
      {High Redshift Galaxy Populations and their Descendants}
\author[Qi Guo and Simon D. M. White]
       {Qi Guo\thanks{Email: guoqi@mpa-garching.mpg.de},
	Simon D.M. White
        \\     
        Max Planck Institut f\"ur Astrophysik, 
        }
\begin{document}

\date{Accepted  ???? ??. 
      Received  ???? ??; 
      in original form 2008 ???? ??}

\pagerange{\pageref{firstpage}--\pageref{lastpage}} 
\pubyear{200?}

\maketitle

\label{firstpage}
\begin{abstract}
We study model predictions for the abundance and clustering of
high-redshift galaxies, and for the properties of their
descendants. We focus on three high-redshift populations: Lyman break
galaxies at $z\sim 3$ (LBGs), optically selected star-forming galaxies
at $z\sim 2$ (BXs), and distant red galaxies at $z\sim 2$ (DRGs).  We
select model galaxies from mock catalogues using the observationally
defined colour and magnitude criteria.  With plausible dust
assumptions, our galaxy formation model can simultaneously reproduce
the abundances, the redshift distributions and the clustering of all
three observed populations. The star formation rates (SFRs) of model
LBGs and BXs are lower than those quoted for the real samples,
reflecting different initial mass functions and scatter in model dust
properties.  About 85\% of model galaxies selected as DRGs are
star-forming, with SFRs ranging up to $\sim 10^2M_\odot$/yr. Model
LBGs, BXs and DRGs together account for less than half of all star
formation over the range $1.5<z<3.2$; many massive, star-forming
galaxies are predicted to be too heavily obscured to appear in these
populations.  Model BXs have metallicities which agree roughly with
observation, but model LBGs are only slightly more metal-poor, in
disagreement with recent observational results.  The model galaxies
are predominantly disk-dominated. Stellar masses for LBGs and BXs are
typically $\sim 10^{9.9}M_\odot$, and for DRGs are $\sim
10^{10.7}M_\odot$. Only about 30\% of model galaxies with
$M>10^{11}M_\odot$ are classified as LBGs or BXs at the relevant
redshifts, while 65\% are classified as DRGs.  Almost all model LBGs
and BXs are the central galaxies of their dark halos, but about a
quarter of DRGs are satellites.  Half of all LBG descendants at $z=2$
would be identified as BX's, but very few as DRGs.  Clustering
increases with decreasing redshift for descendants of all three
populations, becoming stronger than that of $L^*$ galaxies by $z=0$,
when many have become satellite galaxies and their mean stellar mass
has increased by a factor of 10 for LBGs and BXs, and by a factor of 3
for DRGs.  This growth is dominated by star formation until $z\sim 1$
and thereafter by mergers. Merging is predicted to be more important
for LBG and DRG descendants than for BX descendants. Most LBGs and
DRGs end up as red ellipticals, while BXs have a more varied
fate. 99\% of local galaxies with $M_*>10^{11.5}M_\odot$ are
predicted to have at least one LBG/BX/DRG progenitor, and over 70\%
above $10^{11}M_\odot$

\end{abstract}

\begin{keywords}
   galaxies: star formation -- galaxies: colour  -- galaxies:
evolution--
	cosmology: dark matter	 -- cosmology: large-scale structure
\end{keywords}

\section{Introduction}
\label{sec:intro}
The redshift interval $1<z<3$ is a very important epoch in the history
of galaxy formation. During these several billion years, the star
formation rate per unit comoving volume, the abundance of luminous
quasars and the specific merger rate of galaxies all
reached their peak values. This is when the Hubble sequence of
galaxies was established, and most galactic stars were formed.

The last few decades have seen a remarkable development in the
observational study of high-redshift galaxies. Using strong the Lyman
break feature in the spectrum of star-forming galaxies,
\cite{Steidel96} developed 
colour-colour criteria to select so-called Lyman Break Galaxies (LBGs)
at $z\sim 3$ using deep $U_nGR$ photometry. Extensions of this
technique allowed the selection of large samples of star-forming
galaxies at lower redshifts, specifically $z\sim 2.3$ (BXs) and 
$z\sim 1.7$ (BMs) \citep{Adelberger2004}. Recent observations show that there are
many high-redshift galaxies with rather little rest-frame UV
luminosity which are missed in such optically selected
surveys. \cite{Franx2003} have successfully developed near-infrared
colour criteria to select distant red galaxies (DRGs) at $z\sim 2$
using deep $JK$ photometry. Follow-up observations at a variety of
wavelengths have clarified the physical properties of all these
high-redshift galaxy populations by providing star-formation
rates \citep{Erb06a,Reddy06}, stellar
masses \citep{Shapley05,Erb06b,Kriek2006,Papovich2006}, morphologies
\citep{Abraham1996,Papovich2005,Law2007}, dust
luminosities \citep{Webb2003}, kinematics \citep{Pettini2001,Erb06b}
and clustering estimates \citep{Adelberger2005,Quadri2008}.

Each of these observational samples provides information on a limited
subset of the galaxy population at a specific cosmic epoch, and it is
obviously of interest to understand their relation to the population
as a whole, both at the redshift of observation and at other
redshifts, particularly $z=0$ where our knowledge of galaxies is best.
Within the current standard cosmological paradigm, structure formation
in the gravitationally dominant dark matter distribution can be
simulated reliably and is now quite well understood \citep[e.g.][]{SFW}.
For the purposes of modelling galaxy evolution,
it can usefully be idealised as a distribution of dark matter halos of
``universal'' structure which grow steadily in mass through accretion
and merging. Galaxies form through condensation of gas within this
evolving dark halo population, as first set out by \cite{WR1978}. At any given time their distribution can be well modelled by
populating halos with galaxies according to a simple recipe, and then
adjusting parameters to fit the observed abundance and clustering. The
recipe can be based either on a simplified model of the galaxy
formation process or on fitting
formulae which make no direct reference to the underlying physics
\citep{PS2000,Seljak2000}.  The flexibility of the latter Halo
Occupation Distribution (HOD) approach allows excellent fits to galaxy
luminosity functions and clustering properties at any given epoch (see
Cooray \& Sheth 2002 for a review) but provides no natural way to link
populations at different epochs (see \cite{Conroy2008} for a recent
attempt to remedy this). In addition, detailed physical models are
needed to assess how observational cuts on luminosity and colour
affect the properties of high redshift samples.
 
A more straightforward approach is to follow galaxy formation directly
within the evolving dark matter distribution. This extends the
semi-analytic technique developed in the early 1990's 
\citep{WF1991,KWG1993,Cole1994}, and has made increasingly
sophisticated use of information from $N$-body simulations of cosmic
structure formation
\citep{KNS1997,Kauffmann1999,Springel2001,Nature2005}. 
The idea is to design simple parametrised models for the baryonic
physics, based either on observation or on more detailed simulations
of individual systems, and to implement these recipes in the structure
formation framework provided by a dark matter simulation. This
provides a powerful tool for studying the formation, evolution and
clustering of the galaxy population. It is much less
resource-intensive than simulations involving a more direct treatment
of baryonic processes, and as a result it allows the treatment of
larger volumes and the exploration of a wider range of input
parameters and physics recipes.  Recent improvements in such modelling
have included the move to higher resolution N-body simulations,
enabling use of the substructure information they
provide \citep{Springel2001, Helly2003,Hatton2003,Nature2005,Kang2005}
and the inclusion of additional relevant physics, for example, AGN
feedback \citep{Croton2006, Bower2006}, galactic winds
\citep{Bertone2007} and gas stripping in clusters \citep{Font2008}.  
These techniques have previously been used to study the properties of
LBGs by \cite{Blaizot2004}.

 The semi-analytic models used in this paper are implemented on the
\emph{Millennium Simulation} \citep[][hereafter MS]{Nature2005} and
are a minor modification of those presented in De Lucia \& Blaizot
(2007, hereafter DLB07) and Kitzbichler \& White (2007, hereafter
KW07), which are publicly available at the MS download
site\footnote{http://www.mpa-garching.mpg.de/millennium}. These are
refinements of the model originally published as \cite{Croton2006}.
They have successfully matched a wide range of galaxy properties both
locally and at high redshift, but there are some notable discrepancies
which demonstrate that the description of galaxy formation physics
remains incomplete. Our models (and that of KW07) differ from
the model of DLB07 only through the introduction of a redshift
dependence in the way dust is treated. We construct light-cone surveys
of our models as in KW07, and we select LBGs, BXs, and DRGs
exactly as in observational studies. We then compare observed and
simulated samples in terms of their abundance, their clustering and
their distributions of redshift, colour, metallicity and star
formation rate, finding good agreement in most cases. We move on to
examine model predictions for the relation of these various
populations to each other and to the high-redshift galaxy population
as a whole, and for the properties of their descendants at lower
redshift.

Our paper is organized as follows. In Sec.~\ref{sec:simulation}, we
summarize relevant properties of the \emph{Millennium Simulation}, of the
semi-analytic model and of our light-cone surveys. Sec.~\ref{sec:mock}
deals with the selection of model LBG, BX and DRG samples, and
compares them with observed samples.  Sec.~\ref{sec:result} examines
the relation of these model samples to each other and to the full
high-redshift population, and the properties of their
descendants. We summarize our results in Sec.~\ref{sec:conclusion}.

\section{GALAXY FORMATION}
\label{sec:simulation}
\begin{figure}
\bc
\hspace{-0.6cm}
\resizebox{8.5cm}{!}{\includegraphics{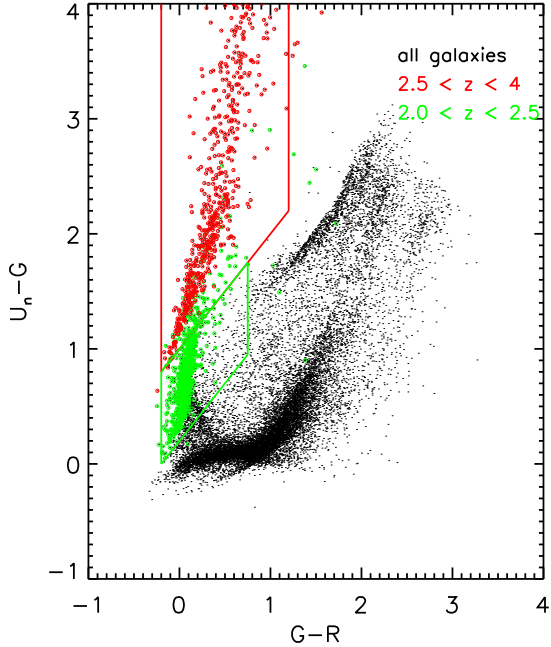}}\\%
\caption{$G-R$ vs. $U_n-G$ diagram for model galaxies selected from
a mock redshift survey. The redshift ranges highlighted by coloured
circles are indicated in the upper right corner. The red and green
boxes outline the original observationally defined selection windows
for LBGs and BXs respectively.  }

\label{fig:colour}
\ec
\end{figure}

\begin{figure}
\bc
\hspace{-0.6cm}
\resizebox{8.5cm}{!}{\includegraphics{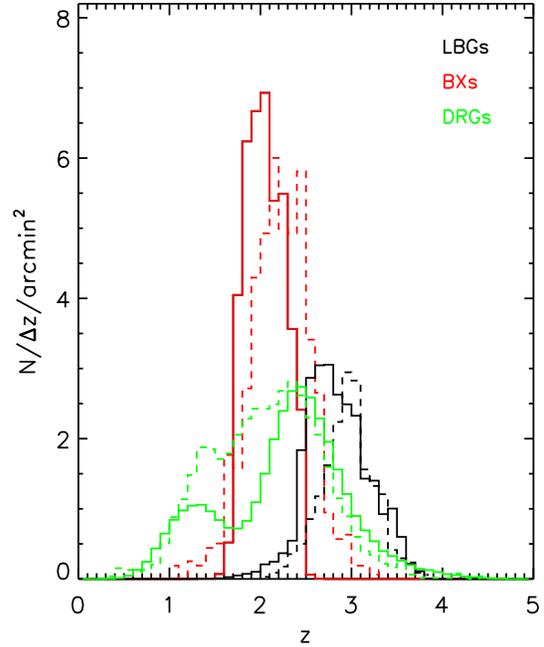}}\\%
\caption{Redshift distributions for LBGs, BXs and DRGs. The solid histograms are
for mock galaxies while dashed ones are for the real observed samples. Black
histograms indicate LBGs, red ones indicate BXs and green ones indicate DRGs
(scaled up in number by a factor of 3 for clarity). Note that the
normalisations are given in terms of the surface density of objects on a
linear scale, and have not been adjusted. The models do indeed reproduce the
observed abundance as a fuction of redshift for all three types of object.}

\label{fig:zdis}
\ec
\end{figure}

\begin{figure}
\bc
\hspace{-0.6cm}
\resizebox{8.5cm}{!}{\includegraphics{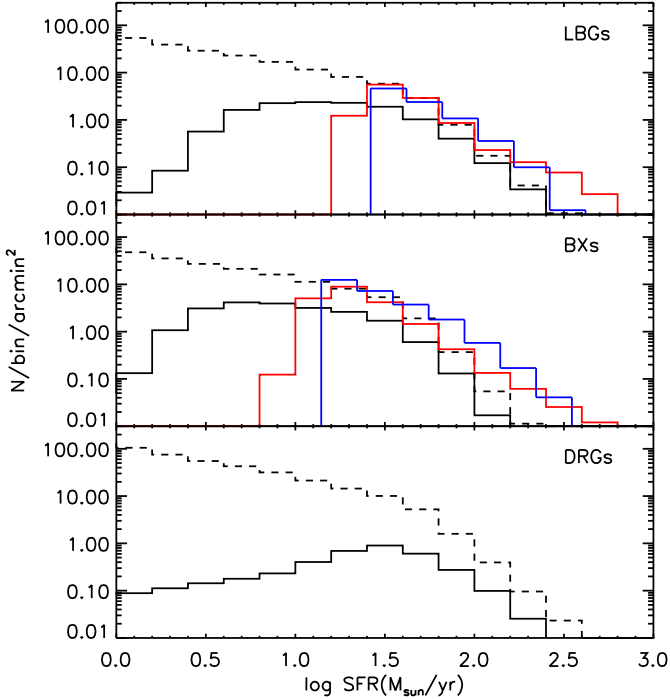}}\\%
\caption{Star formation rate distributions for LBGs, BXs and DRGs. 
Solid black histograms in each panel refer to model galaxies selected
according to the observational colour and magnitude criteria which
define the relevant population. They may be compared with the dashed
black histograms which give results for all model galaxies in the
corresponding redshift range. At all redshifts and SFRs, the
observational criteria select fewer than half of all star-forming
galaxies. Blue histograms in the LBG and BX panels are direct
observational estimates of the SFR distributions for these two
populations taken from Reddy et al. (2008). They are clearly centred at
higher SFR than in the models. The red histograms show what happens if
SFRs are estimated for the model galaxies from their ``observed''
fluxes assuming the same mean correction for extinction and the same
conversion factor from UV luminosity to SFR as
in Reddy et al. (2008). The results agree well with the ``observed''
SFR values but differ substantially from the true values.}

\label{fig:sfr}
\ec
\end{figure}

\begin{figure}
\bc
\hspace{-0.6cm}
\resizebox{8.5cm}{!}{\includegraphics{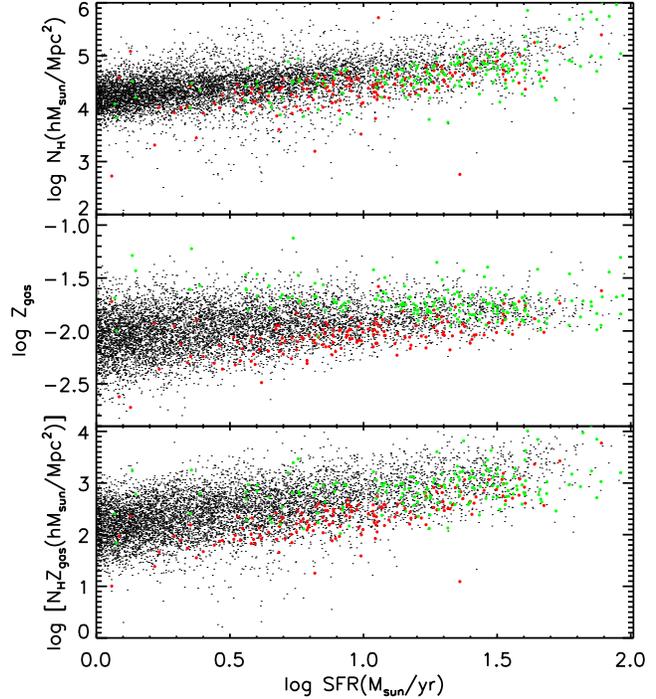}}\\%
\caption{Gas surface density (upper panel), 
metallicity (middle panel) and the product of the two (bottom panel)
plotted against star formation rate for model galaxies at $z=2.2$. Red
and green points are galaxies that satisfy the BX and DRG selection
criteria respectively. Black dots correspond to other galaxies, most
of which are excluded from either sample because they are too faint.}

\label{fig:dust}
\ec
\end{figure}

\begin{figure}
\bc
\hspace{-0.6cm}
\resizebox{8.5cm}{!}{\includegraphics{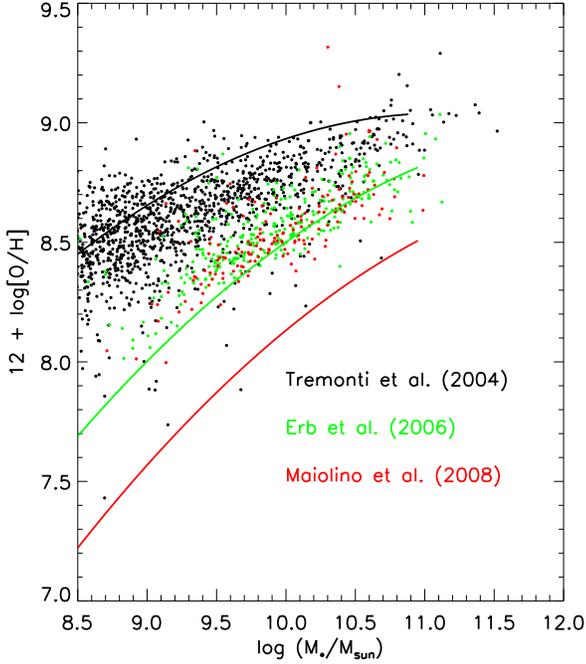}}\\%
\caption{Stellar mass vs. gas-phase  metallicity for
star-forming galaxies. Black dots represent the low-redshift
galaxies, the green dots represent BXs, and the red dots
represent LBGs. Observational estimates of the mean relations between
these quantities are overplotted using curves with the same colour
coding. The sources of the observational relations are indicated
by labels}

\label{fig:metal}
\ec
\end{figure}

\begin{figure}
\bc
\hspace{-0.6cm}
\resizebox{8.5cm}{!}{\includegraphics{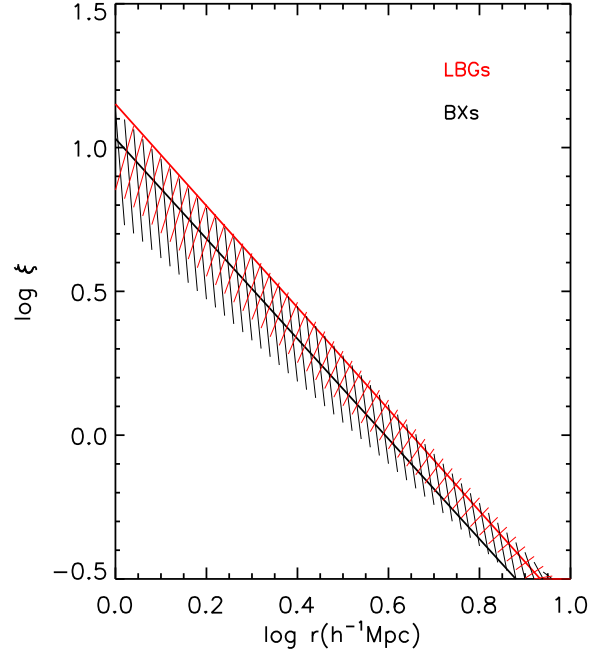}}\\%
\caption{3D correlation functions for LBGs and BXs. The solid 
curves are for galaxies from our mock catalogue, while the hatched regions
indicate the $\pm 1\sigma$ range for the strength of observed
correlations, as quoted by Adelberger et al. (2005). Red curves are for
LBGs and black curves are for BXs.}

\label{fig:mockclu}
\ec
\end{figure}

\begin{figure}
\bc
\hspace{-0.6cm}
\resizebox{8.5cm}{!}{\includegraphics{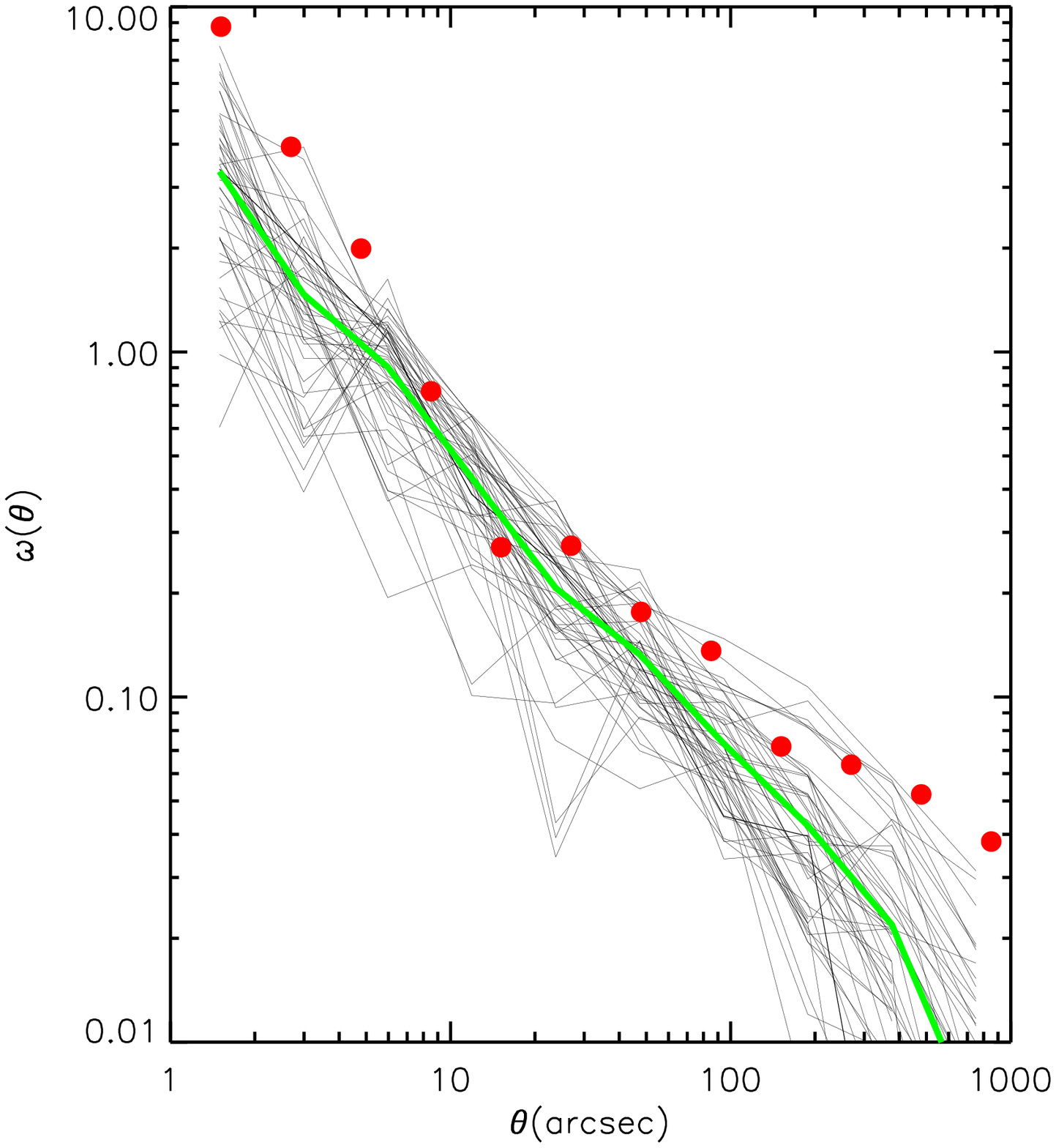}}\\%
\caption{Angular correlation functions for DRGs. Thin black curves are
estimates from 48 independent mock catalogues for fields similar in size to that
analysed by Quadri et al. (2008). The thick green curve is the 
average of these 48 model estimates, while the
red dots indicate the actual measurement of Quadri et al. (2008).}

\label{fig:mockdrg}
\ec
\end{figure}


Our mock galaxy catalogues are constructed by implementing a series of
semi-analytic galaxy formation models on stored merger trees which
represent the entire growth of nonlinear structure, both dark matter
halos and their subhalos, in the very large 
\emph{Millennium Simulation} \citep{Nature2005}. This simulation
assumed a concordance $\Lambda$CDM cosmology with parameters
consistent with a combined analysis of the 2dFGRS and the first-year WMAP data
\citep{Spergel2003}: $\Omega_{\rm m}=0.25$, $\Omega_{\rm b}=0.045$, 
$h=0.73$, $\Omega_\Lambda=0.75$, $n=1$, and $\sigma_8=0.9$, where the
Hubble constant is parameterized as usual as $H_0=
100h$~km~s$^{-1}$Mpc$^{-1}$.  The \emph{Millennium Simulation} represents
the dark matter distribution by following $N=2160^3$ particles from
redshift $z=127$ to $z=0$ in a comoving box of side 500$h^{-1}$
Mpc. Each particle thus has a mass of $8.6\times10^8h^{-1}M_\odot$. The
simulation data were stored at 64 redshifts apaced approximately
logarithmically at early times and linearly at late times. This
spacing determines the time resolution of the semi-analytic modelling
of galaxy formation, A detailed description of the \emph{Millennium
Simulation} can be found in the original article of
\cite{Nature2005} 

The simulation of the evolution of the galaxy population is based on
the modelling techniques developed by the Munich Group. Our galaxy
formation model is almost identical to the publicly available model
of \cite{LB2007} which is itself a refinement of the model originally
implemented on the \emph{Millennium Simulation}
by \cite{Croton2006}. These models include treatments of gas cooling,
star formation and stellar evolution, chemical enrichment, central
black hole formation and growth, material and energy feedback both
from supernovae and from (radio) AGN, and galaxy merging. The reader
is referred to DLB07 and \cite{Croton2006} for detailed descriptions
of how these processes are modelled. As was also the case in KW07, we
have found it necessary to modify the original treatment of dust
attenuation in order to be consistent with high-redshift
observations. We describe this in the next section.

\subsection{Dust Model}

Dust extinction is a crucial ingredient when comparing models of
galaxy evolution with observation.  In this paper we adopt a simple
dust model similar to that used in KW07. The face-on optical depth
is modeled as a function of HI column density and metallicity in the
following way:
\begin{equation}
\tau^Z_{\lambda}=(\frac{A_{\lambda}}{A_V})_{z_{\odot}}\eta_Z(\frac{<N_{H}>}{2.1\times10^{21}cm^{-2}})
\end{equation}
where $A_{\lambda}/{A_V}$ is the extinction curve estimated in
\cite{Cardelli1989} and $<N_{H}>$ is the the average hydrogen column
density. The quantity $\eta_Z=(1+z)^{-0.4}(Z_{gas}/Z_{\odot})^s$ is
the redshift- and metallicity-dependent scaling of dust-to-gas ratio,
where $s=1.35$ for $\lambda<2000 A$ and $s=1.6$ for $\lambda>2000 A$,
The metallicity scalings interpolate between the extinction curves
measured in the Milky Way and in the two Magellanic
Clouds \citep{Guiderdoni1987}. The redshift dependence is
observationally motivated, based on indictions  that
dust-to-gas ratios are lower at high redshift than in the local
universe for galaxies with the same bolometric luminosity and
metallicity \citep{AS2000}. In KW07 a similar model was used, except that the
redshift dependence was set to be $~(1+z)^{-0.5}$ and an artificial
disk size evolution with redshift was assumed so that the equivalent
redshift dependence of dust extinction became $~(1+z)^{-1}$. There is
rather little change in the number count and redshift distribution of
galaxies in our mock catalogues when the power index of the redshift
dependence is changed from -1 to -0.4. We will show later that the
dust model we adopt here produces high-redshift galaxies with the
proper number density both at $z\sim 3$ and at $z\sim 2$.

In addition, our dust model assumes that young stars, defined as stars
younger than $3\times 10^7$yr, are more strongly attenuated than the
rest of the galaxy.  This population is typically still partly
embedded in the molecular clouds from which it formed and so is
more obscured than the general population. Based on the results of
\cite{CF2000}, we assume the mean optical depth in front of young stars to be
three times that which applies to older populations. In addition, we
assign a random inclination to every galaxy and we assume a slab
geometry to calculate the effective extinction from the face-on value.

\subsection{Light-cone}
To make a direct comparison of our galaxy formation simulation with
observation, it is necessary to create a deep light-cone survey which
mimics a real observational survey. Here we use techniques developed
by KW07 to set up the geometry of the light-cone on the periodic
\emph{Millennium Simulation} and to interpolate galaxy properties
from the discrete stored outputs to the continuously varying redshift
coordinate of the light-cone. The only difference to KW07 is in the
dust treatment, as described in the last subsection. We refer the reader
to the original paper for detailed descriptions.

For this paper we construct a mock catalogue on an area of $1.4\times
1.4$ square degrees. We calculate apparent magnitudes in the ($U_n, G,
R$) filters used by the KPNO survey of \cite{Steidel04} and in the
($J, H, K$) filters used by the MUSYC survey
of \cite{Quadri2007a}. IGM absorption is modeled by taking into
account Lyman series line blanketing, continuum absorption by
neutral hydrogen and absorption by heavy elements according to the
recipes of \cite{Madau1995}. We quote all magnitudes in the AB
system. There are a total of 5566388 galaxies in our mock catalogue,
of which 393272 and 224604 galaxies are brighter than apparent
magnitudes of $R<25.5$ and $K<22.86$, respectively. These are the
respective limits of the KPNO and MUSYC surveys.

\section{Mock Catalogue}
\label{sec:mock}

\subsection{Sample Selection}

Owing to the sharp drop-off which it causes at rest-frame 912\AA, the
Lyman break can be used to identify star-forming galaxies at $z\sim3$
from optical broad-band photometry. \cite{Steidel96} developed an
effective criterion for identifying such Lyman Break Galaxies (LBGs)
based on $U_n-G$ and $G-R$ colours as follows:
\begin{eqnarray}
R &\leq& 25.5 \\
U_n-G &\geq& G-R+1.0 
\end{eqnarray}
A similar technique was later developed by \cite{Adelberger2004} to
pick out star-forming galaxies at lower redshifts ($z\sim2$) so-called
BX and BM systems. The criteria proposed for selecting BXs are:
\begin{eqnarray}
R &\leq& 25.5 \\
G-R &\geq& -0.2 \\
U_n-G &\geq& G-R+0.2 \\
G-R &\leq& 0.2(U_n-G)+0.4 \\
U_n-G &\leq& G-R+1.0
\end{eqnarray}
Fig.~\ref{fig:colour} shows a colour-colour diagram for galaxies in our
mock catalogue. Only galaxies with $R$-band apparent magnitude
brighter than 25.5 are shown, in order to mimic the magnitude limit of
the observations. Red circles indicate galaxies in the redshift range
$2.5<z<4.0$, while green circles show galaxies with $2.0<z<2.5$. Black
dots show galaxies at other redshifts.  Red and green boxes outline
the colour selection criteria for LBGs and BXs respectively.  Clearly
our model predicts LBGs and BXs with the proper colours. A careful
comparison with Fig.1 in \cite{Steidel04} shows, however, that the
model LBGs and BXs are about 0.2 mag. bluer in $G-R$ colour than the real
systems. Our colours would be shifted to the red by about 0.1 if we
adopted a Calzetti dust model \citep{Calzetti2000} rather than that
described above. In our mock catalogues the number densities of LBGs
and BXs with $R<25.5$ are 2.37 and 4.35 per square arcminute,
respectively, within 30\% of the observational estimates 
\citep{Steidel03,Steidel04}: 1.8 per sq.arcmin for LBGs and 5.2 per 
sq.arcmin for BXs.

Distant Red Galaxies (DRGs) are K-selected galaxies \citep{Franx2003}
satisfying:
\begin{eqnarray}
K<22.86\\
J-K>1.3
\end{eqnarray}
The number density of such objects in our mock catalogue is 1.16 per
sq.arcmin, in good agreement with the observed density of 1.4 per
sq.arcmin. More impressively, we also reproduce the number density of
a subsample of DRGs with $2<z<3$; there are 0.68 per sq.arcmin. in our
mock catalogue, in excellent agreement with the observed value of 0.66
per sq.arcmin given by \cite{Quadri2008}.
 
\subsection{Redshift Distributions}
We show redshift distributions for LBGs, BXs and DRGs in
Fig.~\ref{fig:zdis}. Black histograms refer to LBGs, red histograms to
BXs and green histograms to DRGs. Solid histograms are for model
galaxies in our mock catalogue, while the dashed histograms are taken
from \cite{Steidel04} for LBGs and BXs and from \cite{Quadri2008} for
DRGs. Note that the normalisations in these histograms have not been
adjusted and that the numbers of galaxies are plotted on a linear
scale. The redshift distributions for model LBGs and BXs are
consistent with observation. although shifted slightly towards lower
redshift. Rather than the observed ranges, $2.7<z<3.4$ for LBGs and
$1.9<z<2.7$ for BXs, In our model the LBGs lie primarily in the range
$2.5<z<3.2$ and the BXs in the range $1.7<z<2.5$. These lower redshift
BXs correspond to the black dots in Fig.~\ref{fig:colour} lying within
the green selection window. The redshift distribution of DRGs is
similar that shown in \cite{Quadri2007a} with a gap between
$1.6<z<2$. However, recent studies \citep{Quadri2008,Grazian2006} show
a continous distribution of redshift over $1<z<3$. Further
observations are needed to reduce photometric redshift uncertainties
and to test whether this gap is an artifact or a real feature. Our
model suggests that it could be real.

In our mock catalogue, contamination by low redshift interlopers is
$\sim0.9\%$ for LBGs ($z<2$), $\sim8\%$ for BXs ($z<1$) and $\sim24\%$
for DRGs ($z<1.8$). These numbers are quite similar to the
observational results: $\sim0.5\%$ for LBGs \citep{Steidel03}, 6\% for
BXs \citep{Steidel04} and $\sim15\%$ for DRGs \citep{Reddy08}. The
fraction of galaxies with redshift $2.5<z<3.2$ and  apparent
magnitude $R<25.5$ which satisfy the LBG colour criteria is 96\% in our
model. The fraction of galaxies with redshift $1.7<z<2.5$ and $R<25.5$
which satisfy the BX colour criteria is 76\%. Both are higher than the
observational values quoted by \cite{Reddy08}: 47\% for LBGs and 58\%
for BXs. Photometric errors, which scatter intrinsically more (or
less) luminous galaxies into (or out of) the selection windows may
partly account for the low observational completeness. On the other
hand, in our model, neither AGN luminosity nor Ly$\alpha$ line
luminosity has been taken into account, both of which may affect
the selection efficiency.

\subsection{Star Formation Rate} 

 We plot the star formation rate distributions of LBGs, BXs and DRGs
in Fig.~\ref{fig:sfr}. Our model galaxies (solid black histograms)
cover a wide range in star formation rate, with average values of
21$M_\odot$/yr for LBGs, 11$M_\odot$/yr for BXs and 24$M_\odot$/yr for
DRGs. Observational estimates of star formation rates are shown by the
blue histograms \citep{Reddy08} and are typically several tens
of $M_\odot$/yr for LBGs and more than $15M_\odot$/yr for BXs,
significantly larger than the values for our model galaxies.

The star formation rate estimates in Fig.~\ref{fig:sfr} were derived
from dust-corrected UV magnitudes. Rather than using the actual SFR
values for our model galaxies, we can estimate ``observational''
values from the apparent magnitudes and colours by applying the
procedures proposed by \cite{Reddy08}. We derive an ``unextincted'' UV
luminosity for each object by combining its $G$ and $R$ magnitudes to
get an approximate rest-frame 1700\AA\ apparent magnitude. We use its
redshift to get the corresponding absolute magnitude. We then multiply
by 4.5 as a mean correction for extinction.  The resulting UV
luminosity is converted to a star formation rate using the relation
given by \cite{Kennicutt1998} for a Salpeter Initial Mass Function
(IMF).

The result is shown in Fig.~\ref{fig:sfr} by the red histograms. These
now agree quite well with the ``observed'' data from \cite{Reddy08},
reflecting the fact that the magnitudes, colours and redshifts of the
model galaxies agree quite well with those of the observed
populations. The distributions of these ``observational'' SFR
estimates disagree badly, however, with the true SFR distributions in
the models. Only a small part of this is due to the fact that the mean
extinction in our model is a factor of 3.9, slightly lower than the
factor of 4.5 proposed in \cite{Reddy08}. Almost a factor of two comes
from the fact that the conversion from UV luminosity to SFR assumes a
Salpeter IMF, while the model assumes a Chabrier IMF. The substantial
difference in shape reflects the fact that extinction factors vary
dramatically from one object to another and are poorly represented by
a mean value. Additional dispersion comes from the finite width of the
galaxy redshift distribution which affects the conversion from
observed magnitudes to rest-frame 1700\AA\ magnitude.

Thus the ``observable'' properties of our mock samples agree quite
well with the real data, but their physical properties suggest that
simple SFR estimates based on mean estimates of obscuration can lead
to substantial systematic and random errors, in particular to an
overestimate of the mean star formation rate.

Comparing the SFR distributions of our photometrically selected
samples of model high-redshift galaxies to that for the high-redshift
population as a whole (dashed histograms) we find that only around
30\% of the galaxies with SFR greater than $5M_\odot$/yr are selected
as LBGs or BXs at $z\sim3$ and $z\sim2.2$, respectively. We show in
Fig.~\ref{fig:dust} scatter plots of gas mass, gas-phase metallicity
and the product of the two against SFR for all galaxies at $z=2.2$
separated into those which satisfy the observational BX selection
criteria (red) and those that do not (black). We also indicate in
green the\ objects which satisy the DRG selection criteria. BX systems
are clearly less obscured than other galaxies with the same SFR. 
Gas content is the the main contributor to this effect, although
metallicity also plays a role. Similar effects are found for model
LBGs -- there are many objects at the same redshift with similar SFR
which are not included in the LBG sample because obscuration makes
them too faint.  It is interesting to see that in the model there are
many DRGs among the highly extincted part of the star-forming galaxy
population at $z=2.2$.
  
The last panel in Fig.~\ref{fig:sfr} is the SFR distribution of
model DRGs. Although selected as red galaxies, most DRGs are, in fact,
star-forming. 85\% of them fill a rather flat SFR distribution
extending from 1 to 40 $M_\odot$/yr and beyond. Only around 15\% of
the DRGs are passive galaxies with SFR less than 1 $M_\odot$/yr. One
thing of interest is to explore the relation between DRGs and BXs. As
can be inferred also from Fig.~\ref{fig:dust}, very few DRGs are also
BXs. We will discuss this further below.

\subsection{Mass-Metallicity Relations} 

Fig.~\ref{fig:metal} plots stellar mass against gas-phase metallicity
for LBGs (red), for BXs (green) and for $z=0$ star-forming galaxies
(black) in our model. Mean observational relations taken
from \cite{Tremonti04} (for local galaxies),from \cite{Erb06b} (for
BXs) and from \cite{Maiolino} (for LBGs) are overplotted as solid
curves of the corresponding colour. The mass-metallicity relations for
local star-forming galaxies and for BXs are moderately well reproduced by the
model. The strong evolution between these two populations reflects the
different physical properties of star-forming galaxies at these two
well-separated epochs and appears somewhat smaller in the model than in the real data. At early times, the galaxies are more gas-rich
and their gas metallicity is relatively low compared to local
galaxies of the same stellar mass. The metallicity predicted for BXs is
higher than than that for LBGs, but the difference is small, since both BXs
and LBGs are selected as UV-bright star-forming galaxies and are
separated by only 1~Gyr. This is inconsistent with the results of
\cite{Maiolino}; the slope of their observed LBG mass-metallicity 
relation is consistent with our model, but its normalisation is much
lower, indicating a strong apparent evolution between $z\sim 2.2$ and
$z\sim3$. This may reflect calibration uncertainties in the differing
metallicity indicators used, rather than true evolution. Further
observational study is needed to confirm that evolution is as strong
as these datasets indicate.

\subsection{Correlation Functions} 

 We show 3D two-point spatial correlation functions for model LBGs
(red) and BXs (black) in Fig.~\ref{fig:mockclu} and we compare them
with observational data from \cite{Adelberger2005}. The solid curves
are the mean functions for model LBGs and BXs, while the hatched
regions between the dashed lines show the $\pm 1\sigma$ range
estimated for their observed clustering. For LBGs the model results
lie within this one sigma band but are near its upper limit. For BXs
the predicted clustering strength is well centred in the observational
band, but the slope of the predicted correlation function is slightly
higher than observed.

The number density of DRGs is quite small. To get a more statistically secure
result and to estimate how cosmic variance may affect observational clustering
measurements, we constructed 48 light-cones for areas of size $0.8\times 0.8$
deg$^2$. In an attempt to better mimic observational uncertainties, we assign
a ``photometric redshift'' to each model galaxy by adding a random
perturbation to its true redshift. Based on the data of \cite{Quadri2008}, we
set the {\it rms} value of this perturbation to be $(1+z)*0.06\sim0.18$.  We
then study the angular correlation function of DRGs with ``photometric''
redshifts in the range $2<z<3$.The results of this exercise are shown in
Fig.~\ref{fig:mockdrg}. The thick green curve represents the mean angular
correlation function and the 48 thin curves represent the angular correlation
functions estimated in the individual $0.8\times 0.8$ deg$^2$ regions.  The
red dots are the observational result of \cite{Quadri2008}. At scales between
7 and 100 arcsec, our predictions overlap the observations quite well. At
larger and smaller scales, they are low, although one or two of the mock
surveys still come close to the observational data.

\cite{Quadri2008} fit line-of-sight projections of simple non-evolving 
double power-law models to the angular correlation data shown in
Fig.~\ref{fig:mockdrg} and concluded that a comoving correlation length of 10.6$\pm$1.6$h^{-1}$Mpc
is needed to match the observations. They noted that in the concordance
$\Lambda$CDM cosmology, this is significantly larger than expected for {\it
any} galaxy population with the observed abundance of DRG's. We confirm this
for own specific model: at $z=2.24$ objects which satisfy the photometric
criteria to be considered DRGs have a comoving correlation length (defined as
the scale at which the 3-D spatial correlation function is unity) of 6.6$h^{-1}$Mpc,
significantly smaller than the \cite{Quadri2008} value. We have tested whether
this apparent discrepancy could be due to the difference in redshift
distributions between model and observed DRGs (see Fig.~\ref{fig:zdis}). We 
increase the abundance of DRGs in the redshift range $1.6<z<2.$ by shifting
the $J-K$ colour cut to 1.1 over this interval. This results in a total number
density of 1.4 DRGs per sq.arcmin and a continuous redshift distribution over
$1<z<3$ with no gap. Thus it reproduces the observed distributions
well. However, model DRGs with photometric redshifts in $2<z<3$ show almost
identical angular correlations to our original samples. We conclude that our
failure to match precisely the observed redshift distributions has no
significant effect on the predicted angular correlations.

It appears that the difference in conclusions here and in \cite{Quadri2008}
stems from the fact that we emphasise the agreement of our model with the
observed angular correlations over the central part of the measured range, and
give less weight to disagreements on the largest and smallest scales,
while \cite{Quadri2008} fit the shape of their measured angular correlations
quite precisely and then use the correlation length as a measure of clustering
strength. At $z=2.24$, a comoving scale of 10.6$h^{-1}$Mpc corresponds to 540 arcsec,
and so is in the range where the observed angular correlations are well above
almost all of our mock catalogues. As a result, the DRG correlation length
quoted by \cite{Quadri2008} is larger than that for our model DRGs. Given the
noise expected for fields of the observed size (illustrated by the scatter
among the thin lines in Fig.~\ref{fig:mockdrg}) it seems likely that a final
resolution of this issue will need to wait for a survey of a substantially
larger area.


\section{The Descendants of  High  Redshift Galaxies}

In the last section we showed that our galaxy formation simulation
reproduces most of the observational properties of the LBGs, BXs and
DRGs. The model may therefore provide a useful guide to the physical
properties of these systems, as well as to their relations to each
other and to lower redshift galaxy populations. We define the
descendant of a high-redshift galaxy population, for example the LBGs,
to be the set of all galaxies at some lower redshift which have at
least one LBG progenitor (which need not be their main progenitor).
In the following, we maximise our statistics by selecting
high-redshift populations from the full Millennium Simulation volume
rather than from a mock catalogue.  This can entrain slight
differences with the results above because the full data are stored
only at discrete epochs and we do not interpolate between them.
\label{sec:result}

\subsection{Number Density, Satellite Fraction and Stellar Mass Growth}

Table 1 lists predictions from our simulation for the abundance of
LBGs, BXs, DRGs and their descendants. The left column is the
redshift. The LBG sample is selected as all objects in the MS volume
at $z=3.06$ which would satisfy the observational criteria to be an
LBG, if seen on our past light-cone at this redshift. The LBG
abundances at lower redshift then refer to the descendants of this
population. The BX and DRG populations are similarly selected, but at
$z=2.2$. The abundance of LBG descendants changes very little from
$z\sim 3$ to $z\sim 1$, but decreases by 12\% by $z=0$ as a result of
mergers. The abundance evolution of BXs and DRGs is similar; there is
little change until $z\sim 1$ followed by drops of 10\% and 17\%
to the present day, respectively. 

The fractions of model LBGs, BXs, DRGs and their descendants which are
satellite galaxies (i.e. are no longer the central galaxy of their
dark halo) are shown in Table 2. LBGs and BXs are almost all central
galaxies, while 24\% of DRGs are already satellite galaxies at the
time when they are identified. The satellite fractions among the
descendants increase rapidly and become comparable by $z\sim1$: 45\%
for LBGs, 39\% for BXs and 35\% for DRGs. These fractions evolve
more slowly at later times. At $z\sim 0$ around half of the
descendants are satellite galaxies in all three samples. While DRGs
had the largest satellite fraction when they were identified, their
descendants actually have the lowest satellite fraction at $z=0$.
\begin{table}
 \caption{Comoving number density ($h^3Mpc^{-3}$) of model LBGs, BXs, DRGs
 and their descendants}

\begin{tabular}{||c||c||c||c||} 

\hline
$z$ & $  LBGs  $ & $  BXs  $ & $ DRGs $ \\
\hline
3       & 0.00143    &       & \\
2       & 0.00142   & 0.00372 & 0.00061\\
1       &  0.00138  & 0.00366 & 0.00058 \\
0        & 0.00121 & 0.00332 & 0.00050\\	
\hline
\end{tabular} 

\end{table}

\begin{table}
 \caption{Satellite galaxy fraction among model LBGs, BXs, DRGs and
 their descendants}

\begin{tabular}{||c||c||c||c||} 

\hline
$z$ & $  LBGs  $ & $  BXs  $ & $ DRGs  $ \\
\hline
3       & 0.031    &       & \\
2       & 0.235   &  0.035 & 0.235 \\
1       &  0.446  & 0.386  & 0.346 \\
0        & 0.484 & 0.490   & 0.386 \\
\hline
\end{tabular} 

\end{table}

\begin{figure}
\bc
\hspace{-0.6cm}
\resizebox{8.5cm}{!}{\includegraphics{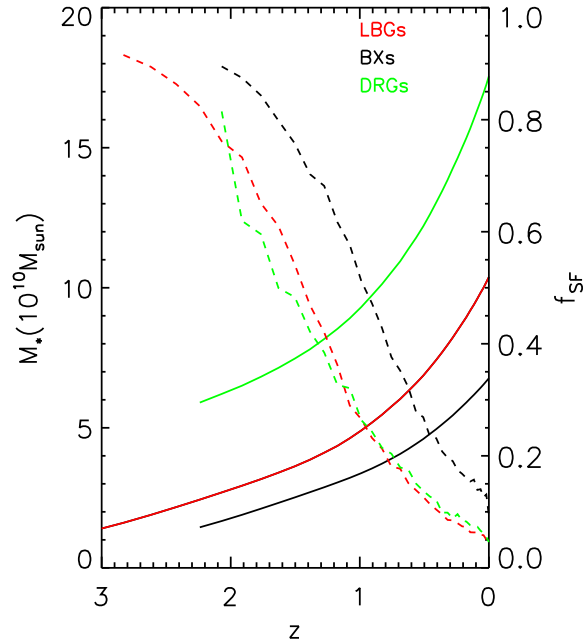}}\\%
\caption{Stellar mass growth in the descendants of high-redshift 
galaxies. The left axis refers to the mean stellar masses of LBGs,
BXs, DRGs and their descendants which are indicated by solid
curves. The right axis refers to the fraction of the mass growth rate
at each redshift which is due to star formation (the dashed
curves). The x-axis is redshift. Red curves are for LBGs, black curves
for BXs and green curves for DRGs.}
\label{fig:macc}
\ec
\end{figure}

Fig.~\ref{fig:macc} shows how the descendants of high-redshift
galaxies grow in mass.. The solid curves indicate the evolution of the
mean stellar mass of each population as a function of redshift. The
red, black and green curves are for LBGs, BXs and DRGs,
respectively. In all three cases the stellar mass increases relatively
slowly until $z\sim 1$ and more rapidly thereafter. This reflects
quiet star-formation-dominated growth at early times, followed by
merger-dominated growth after $z\sim 1$. To see this more clearly, we
also plot as dashed curves the fraction of the mass growth rate at
each redshift which is due to star formation,
$f_{SF}=\frac{\dot{M}_{SF}}{\dot{M}_*}$.  Here $\dot{M}_{SF}$ is the
mean star formation rate while $\dot{M}_*$ is the mean of the total
stellar mass growth rate. At $z\sim 3$, almost 90\% of the stellar mass
growth in LBGs is due to star formation. This fraction drops with time,
and mergers becomes comparable to star formation at $z\sim 1.5$. At
the present day, only 5\% of the growth in stellar mass of LBG
descendants is due to star formation. BXs and their descendants behave
in a similar way, star formation accounts for 90\% of their stellar
mass growth at the time they are identified, but this drops to around
50\% by $z\sim 1$ and to only 10\% at $z\sim 0$. At any given time,
the effect of star formation is more important for BXs than for
LBGs. Interestingly, the evolution of $f_{SF}$ for DRGs is always
close to that for LBGs. Stellar mass growth is dominated by star
formation before $z\sim 1.5$ and by mergers thereafter. The long
steady star-formation-dominated epoch between $z~\sim 2.3$ and $z\sim
1.5$ reflects the fact that most DRGs in our simulation are highly
obscured star-forming galaxies. Note that star formation in this
figure includes the starburst mode during mergers. Gas-rich mergers
can be an important growth mechanism even before $z\sim 1$.

\subsection{Morphology}

\begin{figure}
\bc
\hspace{-0.6cm}
\resizebox{8.5cm}{!}{\includegraphics{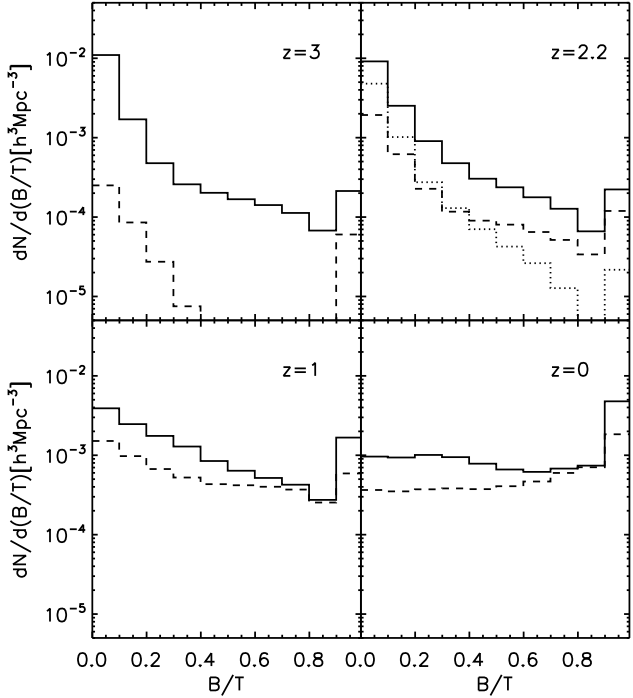}}\\%
\caption{Distributions of bulge-to-total stellar mass ratio for model LBGs
and their descendants. The corresponding redshifts are indicated on
the upper right corner of each panel. The solid histograms are for all
LBGs and their descendants, while dashed histograms show the fraction
of each population which are satellite galaxies. The dotted histogram
in the upper right panel represents the LBG descandents which would be
classified as BXs at $z\sim2.2$.}

\label{fig:lbgmor}
\ec
\end{figure}

\begin{figure*}
\bc
\hspace{-0.6cm}
\resizebox{14cm}{!}{\includegraphics{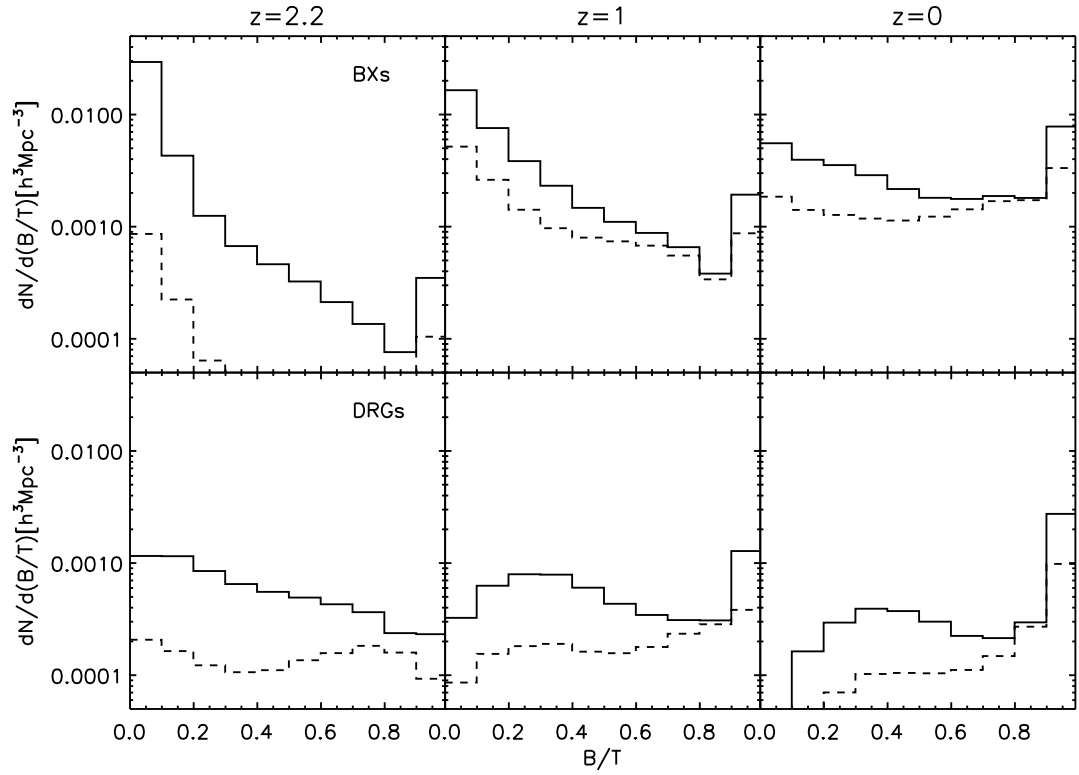}}\\%
\caption{Distributions of bulge-to-total stellar mass ratio for  BXs, 
DRGs and their descendants. The line-style coding of the histograms is
the same as in Fig.~\ref{fig:lbgmor}. The corresponding redshifts are
indicated at the top of each column.}

\label{fig:hzgsmor}
\ec
\end{figure*}

Fig.~\ref{fig:lbgmor} shows distributions of bulge-to-total stellar
mass ratio (B/T) for model LBGs and their descendants. When
identified, most LBGs are disk-dominated (B/T$<$0.5) systems. The
fraction of disk-dominated LBGs remains almost unchanged until $z\sim
2$ and then decreases slowly to $z\sim 1$, showing that gas-rich
major mergers are not a dominant mechanism. (The remnants of major
mergers are assumed to be spheroids in our model.) This fraction
decreases dramatically later on, and more than half of the descendants
are bulge-dominated at $z\sim 0$. Indeed, more than a third 
are ellipticals with B/T greater than 98\%. Given the small fraction
of the $z\sim 0$ growth rate contributed by star formation (less than
10\%) it is clear that the model predicts  dry mergers to dominate
the recent evolution of LBG descendants.  The morphologies of
satellite LBGs and satellite LBG-descendants are similar to those
of their parent samples, gradually changing from disk-dominated to 
spheroid-dominated as the population ages.

The distributions of bulge-to-total stellar mass ratio for model BXs,
DRGs and their descendants are shown in Fig.~\ref{fig:hzgsmor}. The
BXs behave in a very similar way to the LBGs except that the fraction
of disk-dominated systems is larger at all redshifts. Even at $z\sim
0$, around half of all BX descendants are disk-dominated.  At the time
of identification, most model DRGs are also disk-dominated. By $z\sim
1$ the fraction of pure disk systems among their descendants has
dropped by a factor of two. By $z\sim 0$, ellipticals dominate the
population of DRG descendants, accounting for 54\% of the population;
almost no pure disk descendants remain. Mergers clearly play a more
important role in the evolution of DRG descendants than in the evolution of LBG
and BX descendants.
	

\subsection{Stellar Mass Functions}
\begin{figure}
\bc
\hspace{-0.6cm}
\resizebox{8.5cm}{!}{\includegraphics{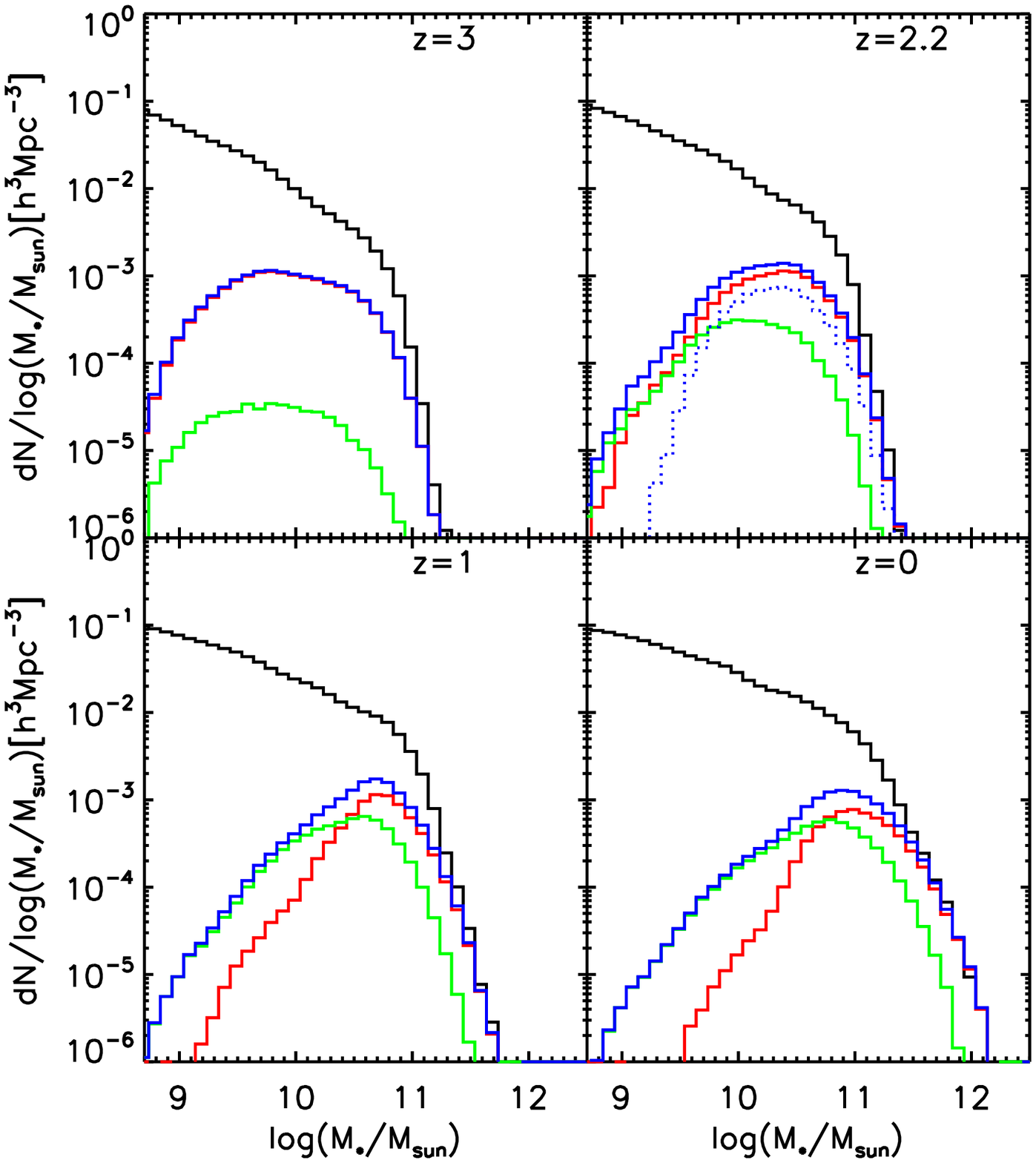}}\\%
\caption{Stellar mass functions for model LBGs and their descendants. Redshifts 
are indicated in the upper right corner of each panel. Blue histograms
are for LBGs and their descendants, red and green histograms split
these populations into central and satellite galaxies,
respectively. For comparison, stellar mass functions for the galaxy
population as a whole are overplotted using black histograms. The
dotted blue histogram in the upper right panel shows those LBG
descendants which would be identified as BXs at $z\sim2.2$ }

\label{fig:lbgmf}
\ec
\end{figure}
\begin{figure*}
\bc
\hspace{-0.6cm}
\resizebox{14cm}{!}{\includegraphics{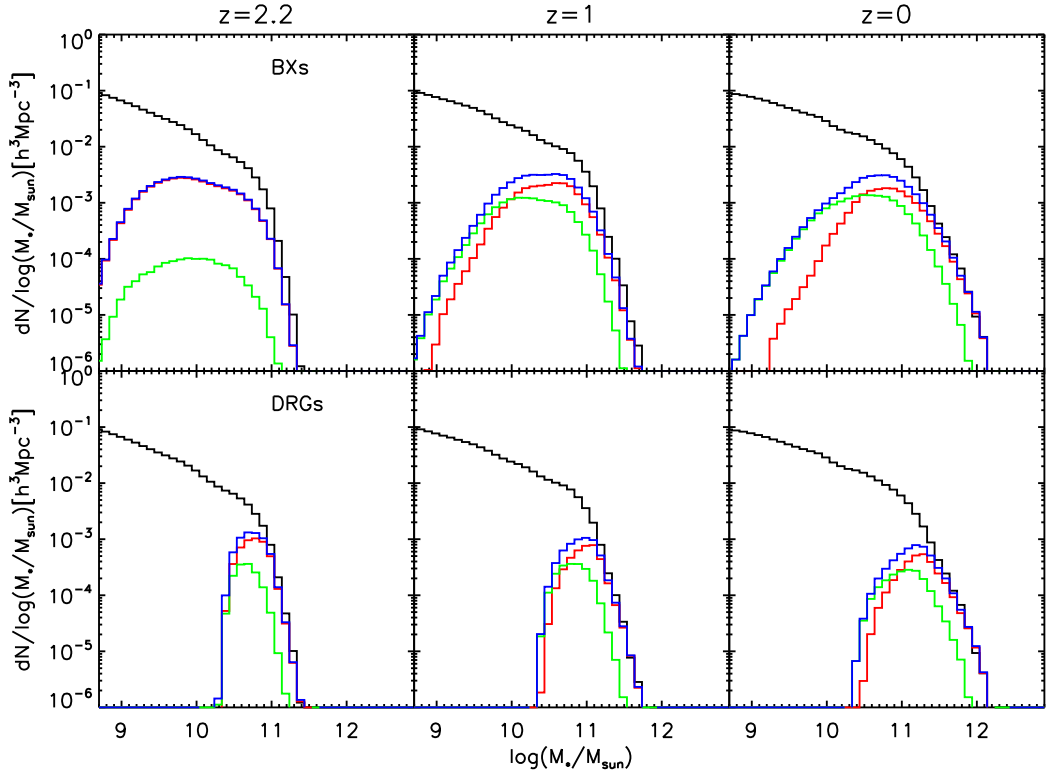}}\\
\caption{Stellar mass functions for model BXs, DRGs and their descendants. 
The histograms use the same colour coding as in
Fig. ~\ref{fig:lbgmf}.}

\label{fig:hzgsmf}
\ec
\end{figure*}

In Fig.~\ref{fig:lbgmf} we plot stellar mass functions for model LBGs
and their descendants and compare these with stellar mass functions
for the galaxy population as a whole at each redshift.  The LBGs cover
a wide range in stellar mass, more than two orders of magnitude, with
a peak at $~10^{9.9}M_\odot$. As time goes by, the median stellar mass
grows by an order of magnitude, becoming more massive than $M^*$ at
$z\sim0$. Very few LBGs end up as central galaxies less massive than
$10^{9.9}M_\odot$. Comparing to the overall stellar mass function at
$z\sim 0$, we find that 88\% of the most massive galaxies
($M_*>10^{11.5}M_\odot$) are LBG descendants. For
$M_*>10^{11.0}M_\odot$ and $M_*>10^{10.5}M_\odot$ the corresponding
fractions are 34\% and 15\%, respectively, Thus, most very massive
galaxies have at least one LBG as their progenitor. Since fewer than
30\% of $z\sim3$ galaxies with $M_*>10^{11}M_\odot$ are identified as
LBGs, many LBGs were accreted onto more massive non-LBG galaxies
during their evolution to low redshift.
  
The stellar mass functions of satellite galaxy LBGs and central galaxy
LBGs are quite similar, but satellite and central galaxies have rather
different mass distributions in the descendant populations. The peak
of the distribution for satellite galaxies is about a factor of two
below that for central galaxies by $z=0$. Satellite descendants grow
less rapidly because almost all mergers occur onto central galaxies
and gas cools only onto central galaxies in our model. As a result
there are also many more satellites in the low-mass tail of the
descendant population.

Stellar mass functions for model BXs, DRGs and their descendants are
shown in Fig.~\ref{fig:hzgsmf}. The results for BXs are again quite
similar to those for LBGs and the mass functions peak at the same
value. The stellar mass of BX descendants is shifted to slightly lower
values than that of LBG descendants and peaks at $10^{10.7}M_\odot$ at
$z\sim0$. At this time 86\% of galaxies with $M_*>10^{11.5}M_\odot$
are BX descendants and 50\% of galaxies with $M_*>10^{11}M_\odot$. For
comparison, only 30\% of $z\sim 2$ galaxies with $M_*>10^{11}M_\odot$
are classified as BXs. Unlike the LBGs and BXs, model DRGs span a
narrow range of stellar mass at the time they are identified, peaking
at $10^{10.7}M_\odot$. By $z\sim 0$ the typical mass of their
descendants has increased only by a factor of 3, but almost no
descendant is less massive than the Milky Way. DRGs account for more
than 65\% of galaxies more massive than $10^{11}M_\odot$ at $z\sim
2.2$ and their descendants at $z=0$ account for more than 84\% of
galaxies more massive than $10^{11.5}M_\odot$.  Many LBGs and BXs are
accreted onto massive DRG descendants by $z\sim0$.
\subsection{Colour-Stellar Mass Distributions}

\begin{figure}
\bc
\hspace{-0.6cm}
\resizebox{8.5cm}{!}{\includegraphics{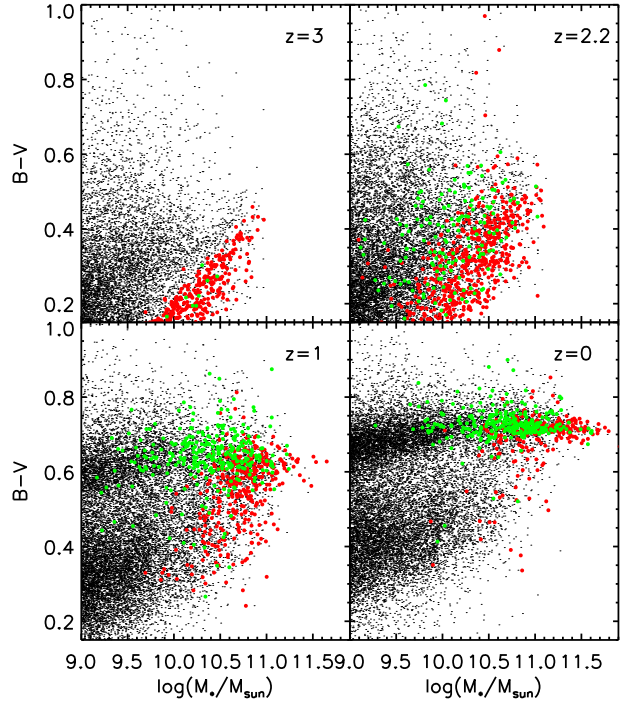}}\\%
\caption{B-V colour vs. stellar mass for model LBGs and their 
descendants at the redshifts indicated in each panel. Red points
denote central LBGs and central LBG descendants, while green points
denote satellite LBGs and satellite descendants. Black dots indicate
other galaxies at the same redshift. }

\label{fig:lbgmc}
\ec
\end{figure}

\begin{figure*}
\bc
\hspace{-0.6cm}
\resizebox{14cm}{!}{\includegraphics{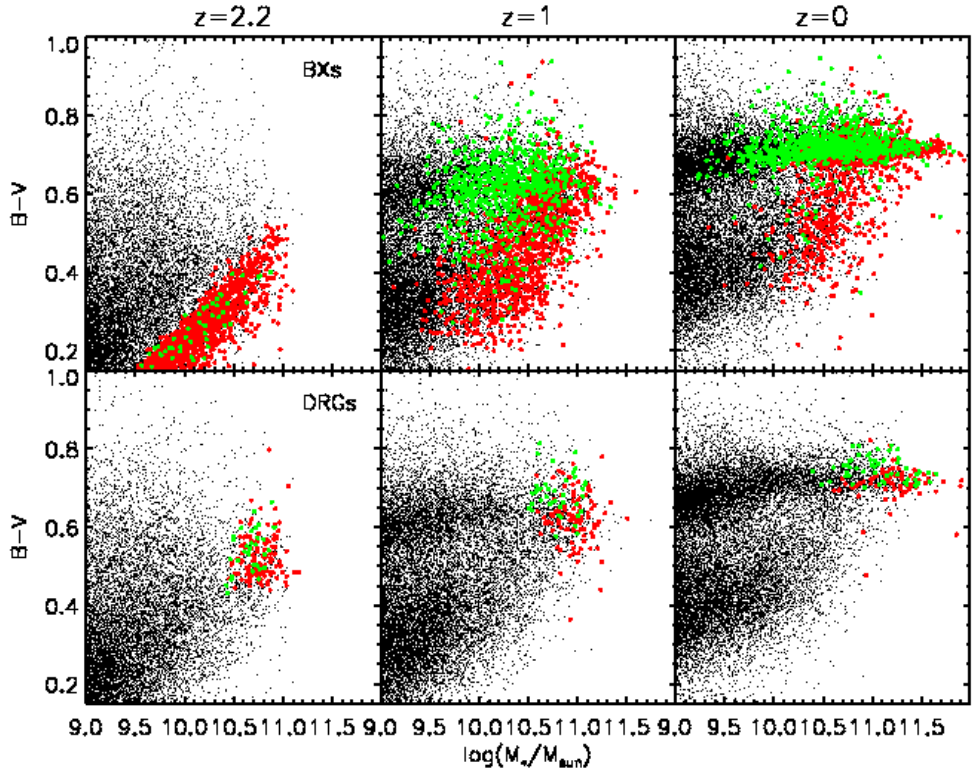}}\\%
\caption{B-V colour vs. stellar mass for model BXs, DRGs and their 
descendants using the same colour coding as Fig.~\ref{fig:lbgmc}.}

\label{fig:hzgsmc}
\ec
\end{figure*}

In Fig.~\ref{fig:lbgmc} we show a scatter plot of rest frame $B-V$
colour (AB system) against stellar mass for model LBGs and their
descendants, as well as for other galaxies at the corresponding
redshifts. When identified, the LBGs are mostly blue galaxies,
although none are part of the small red population which already exists at $z\sim3$. The
blue fraction decreases with time. Although their stellar mass has
increased by a factor of 2 by $z\sim 2.2$, most of the descendants,
especially the central descendants, are blue, indicating a high
specific star formation rate.  At $z\sim 1$ a significant fraction of
the descendants lie on the red sequence. These objects are primarily
satellite galaxies where star formation is suppressed. By $z\sim 0$,
most of the descendants, even central descendants, have moved to the
red sequence.

Fig.~\ref{fig:hzgsmc} shows similar scatter plots of colour against
stellar mass for model BXs, DRGs and their descendants. Yet again the
behaviour of the BXs is very similar to that of the LBGs: When
selected they are almost all blue galaxies which then gradually move
to the red sequence. There are relatively more blue central BX
descendants at $z\sim 0$ than blue central LBG descendants.  Unlike
the LBGs and BXs, most DRGs reside at the massive end of the red
sequence in the stellar mass colour diagram, with a typical B-V colour
of 0.5. As time goes by, the colour of DRGs gets even redder and they
remain on the red sequence. There is almost no difference in colour
between their central and satellite descendants.

\subsection{The Dark Halos of LBGs}

\begin{figure}
\bc
\hspace{-0.6cm}
\resizebox{8.5cm}{!}{\includegraphics{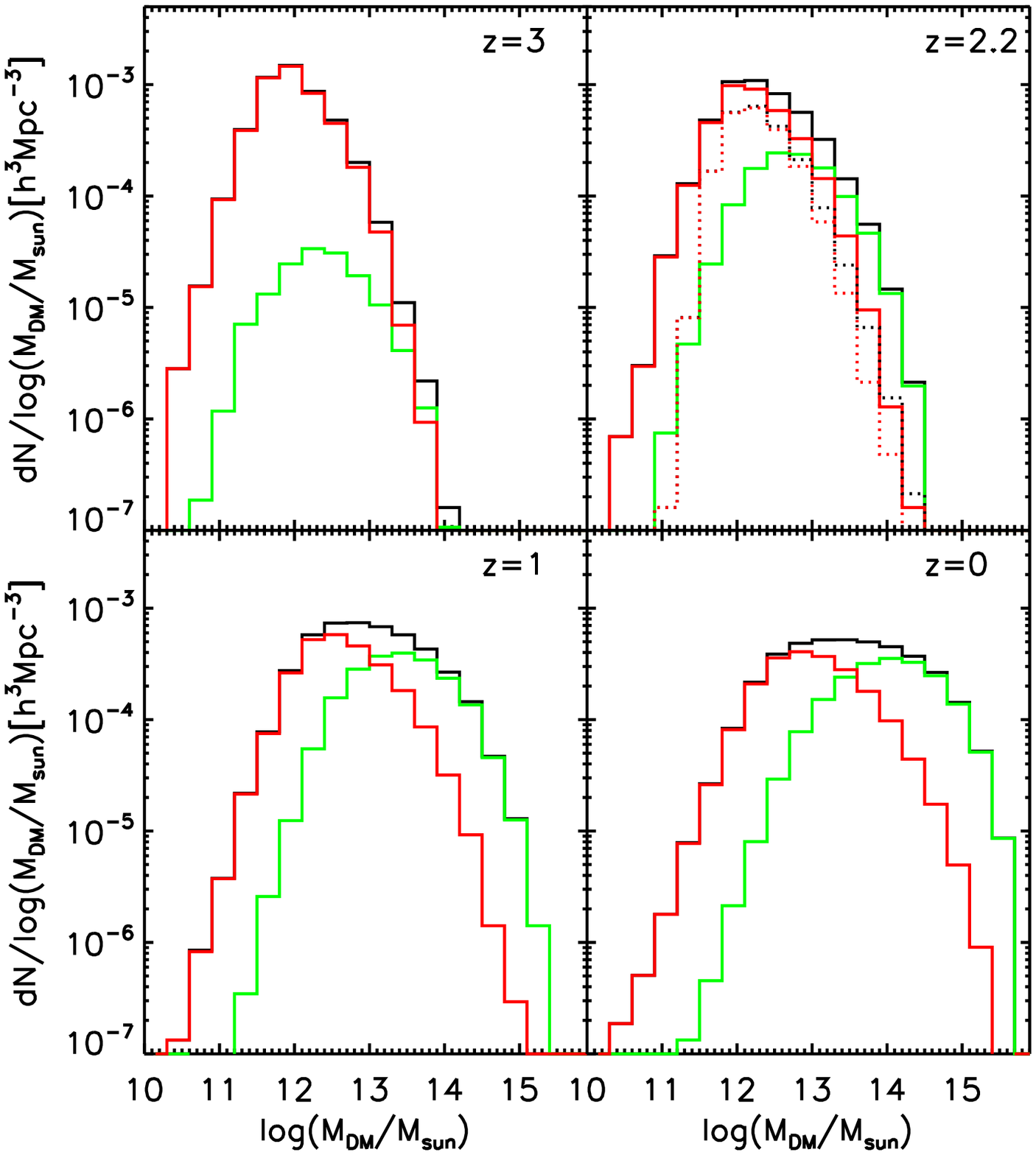}}\\
\caption{Mass distributions for the dark halos of model LBGs
and their descendants. Black histograms are for all objects, while red
and green histograms split the sample into central and satellites
galaxies, respectively.  The dotted black histogram in the upper right
panel shows the LBG descendants which would be classified as BXs,
while the red dotted histogram shows the subset of these objects which
are also central galaxies. Dark halos are multiply counted in these
histograms if they contain more than one LBG or LBG descendant.}

\label{fig:lbgdm}
\ec
\end{figure}

\begin{figure*}
\bc
\hspace{-0.6cm}
\resizebox{14cm}{!}{\includegraphics{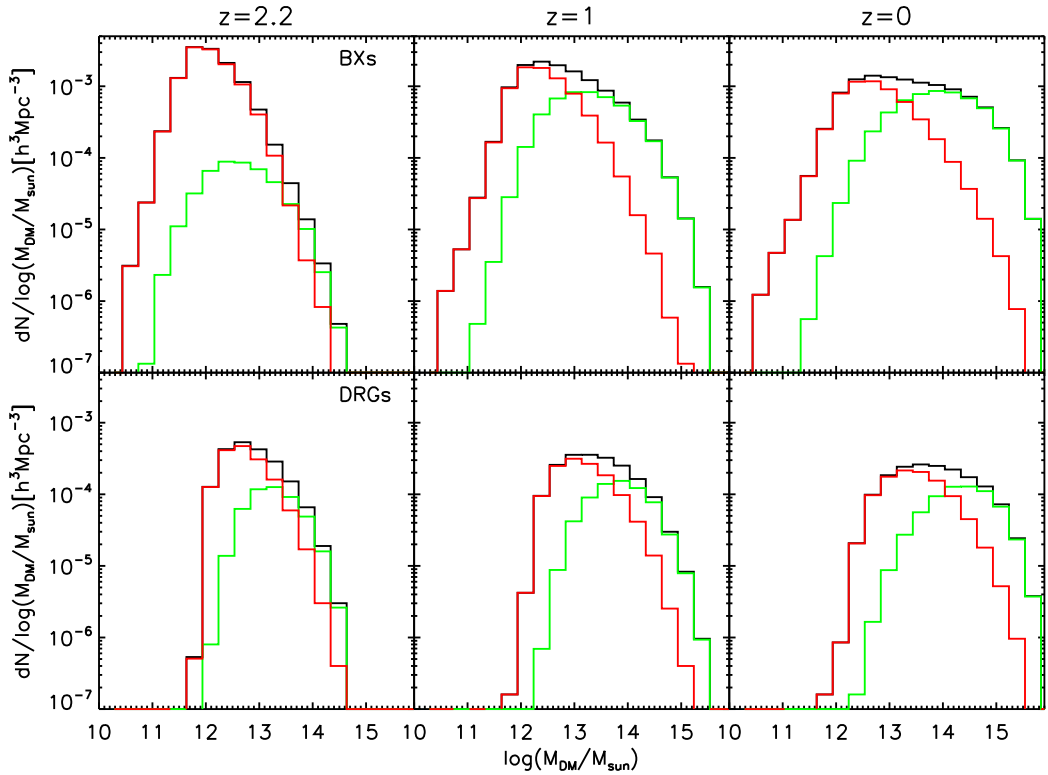}}\\%
\caption{Mass distributions for dark halos of BXs, DRGs and their 
descendants with the same colour coding as in Fig.~\ref{fig:lbgdm}.}

\label{fig:hzgsdm}
\ec
\end{figure*}

Fig.~\ref{fig:lbgdm} shows mass distributions for the dark matter
halos of model LBGs and their descendants. The dark matter halo here
is taken to be the friends-of-friends halo defined by linking together
dark matter particles separated by less than 0.2 of the average
interparticle separation \citep{Davis1985}. Each FOF halo typically
contains several galaxies: one central galaxy and some satellites.
Thus, these histograms may count a given halo more than once if it
contains several LBGs or LBG descendants.  The peak of the halo mass
distribution is at $10^{12}M_\odot$ at $z\sim 3$ and moves to
$10^{13.4}M_\odot$ by $z\sim 0$, corresponding to the mass of a large
galaxy group. At high redshift, central and satellite LBGs live in
halos of similar mass, but at lower redshifts satellite descendants
tend to live in more massive halos than central descendants. In fact,
the typical present-day environment of central LBG descendants is a poor
galaxy group with mass $\sim 10^{13}M_\odot$, while the typical
environment of the satellite descendants is a cluster with mass
$\sim10^{14}M_\odot$.

Fig.~\ref{fig:hzgsdm} shows halo mass distributions of model BXs, DRGs
and their descendants. As before, BXs are distributed in a similar way
to LBGs, although their halos are shifted to noticeably lower masses,
peaking at $10^{11.8}M_\odot$ at $z\sim 2$ and then shifting to
$10^{12.6}M_\odot$ by $z\sim 0$. For DRGs, the halo masses are larger,
peaking at $10^{12.6}M_\odot$ at $\sim2.2$ and shifting to
$10^{13.4}M_\odot$ by $z\sim 0$. Note that the distribution of halo
masses for DRGs is quite narrow compared to that for BXs, consistent
with their narrower distribution in stellar mass
(Fig.~\ref{fig:hzgsmf}). The satellite descendants of BXs and DRGs
also more likely be found in more massive halos than their central
descendants.


\begin{figure}
\bc
\hspace{-0.6cm}
\resizebox{8.5cm}{!}{\includegraphics{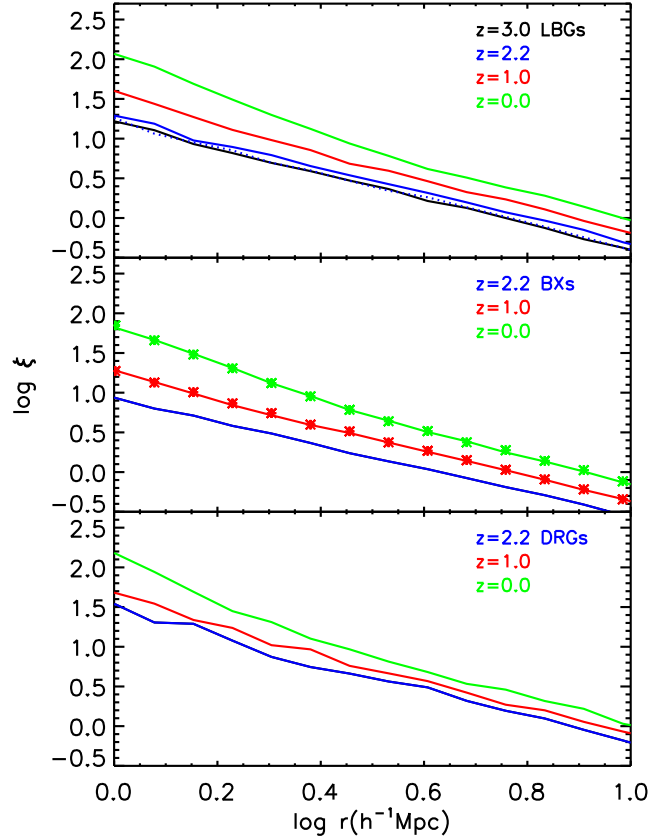}}\\%
\caption{3D correlation functions for LBGs, BXs, DRGs and for their
descendants. Redshifts are indicated at the right upper corner of each
panel using the appropriate colour. Solid curves represent the
results for full samples of LBGs, BXs, DRGs or their descendants. The
dotted curve in the upper panel represents the correlation for LBG
descendants which are BXs at $z\sim2.2$. Stars in the middle panel
show correlations at later times of the dark matter particles
identified as BXs at $z\sim 2.2$ and so represent how correlations
would evolve in a non-merger scenario.}

\label{fig:HzGsclu}
\ec
\end{figure}

\subsection{Descendant Correlations}

In a previous section we showed that the correlation functions of our
model high-redshift galaxies match observation quite well. Here we
look at the evolution of their clustering. In Fig.~\ref{fig:HzGsclu},
we plot 3D correlation functions as a function of redshift.  For all
three populations, the clustering of descendants gets stronger towards
lower redshift. For LBG descendants, the correlation length is
$9.7h^{-1}$Mpc at $z=0$, as estimated from a power law fit over
$3h^{-1}$Mpc$<r<10h^{-1}$Mpc.  This is about twice the comoving value
at $z\sim3$. The difference between $z\sim3$ and $z\sim2.2$ is quite
small compared to changes at later times. At low redshifts there is an
obvious turn-up in clustering strength at small scales, indicating the
increasing importance of satellite galaxies which we already noted
above.

The clustering of BX descendants increases similarly with time, but is
always weaker than that of LBG descendants. Their correlation length
reaches $8.1h^{-1}$Mpc at $z=0$. The turn-up at small scales also
shows up for BX descendants at low redshift. The comoving correlation
length of DRGs evolves less than that of LBGs and BXs, reaching
$10h^{-1}$Mpc at $z=0$. Indeed the correlations of DRG descendants and
LBG descendants are almost identical at $z=0$. For all three
populations, the present-day correlation lengths are larger than for $L^*$ galaxies, as might be expected given that their typical
stellar masses are higher than $M^*$.

It is interesting to check whether galaxy merging has any significant
effect on these descendant correlations. We have tested this by by
tagging the dark matter particles associated with each BX galaxy at
the time it is identified, and then calculating correlations for this
particle set at later redshifts.  The results are indicated by stars
in the central panel of Fig.~\ref{fig:HzGsclu}. Clearly, this
procedure produces results which are indistinguishable from those
obtained by following BX descendants through the \emph{Millennium
Simulation} galaxy trees. This validates one part of the simplified
recipe adopted by \cite{Conroy2008} and \cite{Quadri2007b} when
estimating clustering for the descendants of high redshift galaxy
populations based on simple HOD assignments of the high-redshift
objects to dark halos. Valid results from such recipes will still, of
course, require that they assign galaxies to the correct high-redshift
halos.

\subsection{Relation between LBGs, BXs and DRGs}

At $z\sim 2.2$, only 0.8\% of model BXs and 10\% of LBG descendants are
classified as DRGs. Conversely, 4.7\% and 25\% of DRGs are identified
as BXs and LBG descendants, respectively. The overlap between DRGs
and BXs or LBG descendants is thus quite small.

In contrast, LBGs and BXs are closely correlated. Fully 45\% of LBG
descendants at $z\sim 2.2$ are identified as BXs. We illustrate the
properties of these particular LBG descendants (which we refer to as
LBG-BXs in the following) in the right upper panels of
Fig.~\ref{fig:lbgmor}, Fig.~\ref{fig:lbgmf} and Fig.~\ref{fig:lbgdm}
using dotted histograms. Their distribution in stellar mass is similar
to that of all LBG descendants, but with fewer galaxies in the
low-mass tail, which consists of satellite galaxies and small central
galaxies. Very few satellite galaxies have enough star formation to
qualify as BXs in our model, and only 5\% of LBG-BXs are satellites,
as compared to 23.5\% of all LBG descendants. Some of the more massive
LBG descendants fail to be identified as BXs because they are
rich in gas and heavy elements, and the associated extinction pushes
them outside the BX detection window, even though they are strongly
star-forming.  LBG-BXs are more massive than the overall BX
population at $z\sim2.2$, and this is reflected in their stronger
clustering, as shown in the upper panel of Fig.~\ref{fig:HzGsclu}.
Their clustering is similar to that of LBG descendants in general. A related coincidence shows up in the halo mass
distributions which are very similar for the two populations
(Fig.\ref{fig:lbgdm}). Since mergers play a minor role in the
evolution of LBG descendants before $z\sim2.2$, the LBG-BXs are, as
expected, mainly disk-dominated.


\section{Summary and Discussion}
\label{sec:conclusion}
We have used the \cite{LB2007} model for galaxy formation within the
\emph{Millennium Simulation} to explore the likely physical nature of
observed high-redshift galaxies, specifically Lyman Break Galaxies, BX
galaxies and Distant Red Galaxies, and their likely descendants at
lower redshift. We first built mock catalogues in order to compare
observed high-redshift galaxies with similarly selected objects in the
simulation. We then used the full galaxy catalogue from the simulation
to study the descendants of the high-redshift populations. We found it
necessary to modify the original DLB07 dust model in order to match
the abundance and colour of the observed high-redshift populations,

Impressively, with a proper dust model it is possible to match the observed
abundances, redshift distributions and clustering of all three high-redshift
populations in a model which also fits the properties of low-redshift
galaxies. The descendants of all three populations become more strongly
clustered at lower redshifts, and all are more clustered than $M^*$ galaxies
today. A turn-up in clustering strength at small scales is evident at
$z\sim0$, which reflects the high satellite fraction among the
descendents. The clustering of DRGs is the least consistent with observation.
Angular correlations are well reproduced on scales between 7 and 100 arcsec
but our model appears more weakly clustered than real DRGs on both smaller
and larger scales. This results in an estimated correlation length for the
observed sample which is larger than that of our model DRGs.  Our results show
the expected scatter in angular correlation estimates to be quite large for
areas as small as that currently observed, so a final judgement on this issue
will require significantly larger observational surveys.

Together, model DRGs and BXs account for only 30\% of the strongly
star-forming galaxies ($\dot{M}_*>20M_\odot/yr$) at $z\sim 2$. Most of the rest are gas- and
metal-rich systems which are strongly obscured. In contrast, the model
suggests that only 15\% of all galaxies with $M_*>10^{11}M_\odot$
at $z\sim 2$ (most of which do indeed have high star formation rates) are
missing in current optical and near-infrared surveys. Interestingly,
the model predicts most DRGs to be star-forming galaxies. Their
average SFR is even higher than those of LBGs and BXs, and they
account for more than 65\% of the galaxies with $M_*>10^{11}M_\odot$
at $z\sim2.2$.  There is rather little overlap between these
star-forming DRGs and BXs or LBG descendants. On the other hand, half
of all LBG descendants are identified as BXs at $z\sim2.2$. The
physical properties and the clustering of these LBG-BXs are quite
similar to those of other LBG descendants. These consist of three
classes of galaxy: smaller galaxies where the potential is too shallow
to retain the gas expelled by supernova feedback, satellite galaxies
where the interstellar gas has been exhausted, and gas-rich galaxies
where the dust extinction is strong.

Our simulation roughly reproduces the observed relation between
stellar mass and gas-phase metallicity for local star-forming galaxies
and for BX galaxies, but the gas-phase metallicities predicted for
LBGs, although slightly lower than for BXs, are well above the values
estimated for real LBGs by \cite{Maiolino}.

Most model LBGs, BXs and DRGs are disk-dominated systems, residing at
the centres of their own halos. LBGs and BXs are selected as blue
galaxies and so exclude the significant population of red galaxies
which is already present at these redshifts. Nevertheless, by $z\sim0$
at least half of their descendants are bulge-dominated and red.  The
typical stellar masses of LBGs and BXs increase by an order of magnitude
by $z\sim0$, whereas DRG stellar masses increase by a smaller factor,
$\sim 3$. Star formation is the main driver of growth before $z\sim
1$, then mergers become dominant.  Most low-redshift massive galaxies
($M_*>10^{11}M_\odot$) descend from at least one of these high
redshift populations.  Correspondingly, most descendants are massive
galaxies living in massive dark matter halos today. Many of them are
satellite galaxies in galaxy clusters. The central galaxies in rich
clusters ($M>10^{14.7}M_\odot$) typically have 8.4 LBG, BX or DRG
progenitors, and on average the stars present in the high-redshift
galaxies account for around 50\% of the current stellar mass.  Thus
while the observed high-redshift galaxies have contributed
significantly to today's massive galaxies, it appears that the
relation between the two populations is quite complex.




\section*{Acknowledgments}
We thank Roderik Overzier, Cheng Li, Alice Shapley, Ryan Quadri, Marijn Franx,
 Gabriella De Lucia and Jeremy Blaizot for useful discussion. The
 public databases used in this work are at
 http://www.mpa-garching.mpg.de/millennium. We are grateful to
 Gerard Lemson for help in using these databases.  

\bibliographystyle{mn2e}

\bibliography{LBG}

\end {document}